\documentstyle[12pt]{article}

\newcommand{\bfx}{{\bf x}}
\newcommand{\Schrodinger}{Schr\"{o}dinger\ }

\begin{document}
\begin{titlepage}
\begin{center}

\hfill BU-CCS-970103\\
\hfill PUPT-1678\\
\hfill quant-ph/9701019\\

\vskip .2in

{{\Large \bf Simulating Quantum Mechanics\\ on a Quantum
Computer\footnote{Expanded version of a talk given by WT at the
PhysComp '96 conference, Boston University, Boston, November 1996}}}

\vskip .2in

Bruce M.\ Boghosian\\
{\small \sl Center for Computational Science} \\
{\small \sl Boston University} \\
{\small \sl 3 Cummington Street} \\
{\small \sl Boston MA 02215, USA} \\
{\small \tt bruceb@bu.edu} \\[0.3cm]
\vskip .1in

Washington Taylor IV\\
{\small \sl Department of Physics,} \\
{\small \sl Princeton University,} \\
{\small \sl Princeton, New Jersey 08544, U.S.A.} \\
{\small \tt wati@princeton.edu} \\[0.3cm]
\end{center}
\vskip .2in

\begin{abstract}
Algorithms are described for efficiently simulating quantum mechanical
systems on quantum computers.  A class of algorithms for simulating
the \Schrodinger equation for interacting many-body systems are
presented in some detail.  These algorithms would make it possible to
simulate nonrelativistic quantum systems on a quantum computer with an
exponential speedup compared to simulations on classical computers.
Issues involved in simulating relativistic systems of Dirac and gauge
particles are discussed.
\end{abstract}

\end{titlepage}
\newpage
\renewcommand{\thepage}{\arabic{page}}
\setcounter{page}{1}

\section{Introduction}

There has been a great deal of excitement in the field of quantum
computing over the last few years, due largely to the result of Shor
\cite{Shor} showing that large numbers can be factored on a quantum
computer in a time that scales polynomially with the number of digits.
Shor's work was based on earlier results showing that certain oracle
problems could be solved on a quantum computer exponentially faster
than on a classical computer \cite{oracle}.  Since it is believed that
factoring is an exponentially hard problem on a classical computer,
Shor's result provided for the first time a concrete demonstration
that quantum computers can achieve useful results much more
efficiently than classical computers.  Despite a great deal of work in
this field, however, there has been little progress toward discovering
other algorithms which give the quantum computer an exponential
advantage in performance.

Early in the development of quantum computing, it was suggested by
Feynman \cite{Feynman} that it might be possible to simulate quantum
mechanical systems exponentially faster on quantum computers than on
classical computers.  Feynman argued that understanding this issue
might help to quantify the differences between the computational
capacities of classical and quantum computers.  Recently, it was
argued by Lloyd \cite{Lloyd} that for a fairly general class of
quantum systems, particularly for discrete systems, it should be
possible to achieve an exponential speedup when simulating quantum
mechanical systems on a quantum computer.  In this paper we will
discuss algorithms which are concrete realizations of these general
arguments.

There are three particular types of quantum theories which we would
like to try to simulate on a quantum computer:
\begin{itemize}
\item Nonrelativistic many-body systems (the \Schrodinger equation)
\item Relativistic many-fermion systems (the many-body Dirac equation)
\item Gauge field theories (quantum Yang-Mills theories)
\end{itemize}
In this paper we describe in some detail a class of algorithms for
simulating the many-body Schr\"odinger equation.  These algorithms
were presented from a different point of view in \cite{bw1,bw2}.
Based on simple Quantum Cellular Automata (QCA) \cite{gz} and Quantum
Lattice-Gas Automata (QLGA) \cite{Meyer} models, these algorithms can
be used to simulate systems of interacting nonrelativistic quantum
particles with speedup exponential in the number of particles in the
system.  We give a short discussion of progress which has been made
toward developing algorithms for simulating systems of many Dirac
particles, and we discuss prospects for simulating gauge field
theories on a quantum computer.

\section{Quantum computers}

We begin with a brief description of a universal quantum computer, in
order to fix notation.  For recent reviews of quantum computing in
general, see  \cite{qc}.  The state space of a quantum computer is
defined to be the Hilbert space associated with a finite number ($N$) of
two-state quantum bits (q-bits).  A natural basis for this Hilbert
space is given by the set of $2^N$ states
\begin{equation}
| {\bf \sigma} \rangle = | \sigma_1 \sigma_2 \cdots \sigma_N \rangle
\;\;\;\;\; {\rm  where} \; \sigma_i \in\{\uparrow,\downarrow\}.
\end{equation}
A single unit of computation is defined to be the action of an
arbitrary two-bit gate on the state of the system.  Mathematically,
this corresponds to acting on the state of the system with a
$2^N\times 2^N$ matrix which is a tensor product of an arbitrary
$4\times 4$ unitary matrix (acting on the Hilbert space associated
with a pair of q-bits $\sigma_i, \sigma_j$), with an identity operator
of dimension $2^{N -2} \times 2^{N -2}$.  It has been shown
\cite{universal} that such two-bit gates are universal for quantum
computation.

A measurement in a quantum computer is performed by measuring the
states of some of the q-bits.  It is sufficient to restrict attention
to measurements of the states of individual spins with respect to the
canonical basis.  As dictated by quantum mechanics, a single q-bit in
the quantum state $(\psi_\uparrow |\!\uparrow\, \rangle + \psi_\downarrow
|\!\downarrow \,\rangle)$ will be measured to be in the state $\uparrow$
(or $\downarrow$) with probability $| \psi_\uparrow |^2$ (or $|
\psi_\downarrow|^2$).

The definition of a universal quantum computer is closely modeled on
the conceptual framework of classical computation.  It has been
suggested  \cite{others} that a more general definition of a
quantum computer may be desirable, as there may be physical systems
which exhibit the capacity for performing useful quantum computation
which do not fit into this axiomatic framework.  We will discuss
briefly in the following section an example of a computation which is
difficult on a standard universal quantum computer, which might be
more easily implemented on a generalized quantum computer.  However,
we will otherwise remain within the context of the standard definition
of a universal quantum computer in terms of q-bits and two q-bit
operations.

\section{Simulating the \Schrodinger equation}
In this section we describe a class of algorithms for simulating the
many-body \Schrodinger equation on a quantum computer with exponential
speedup over simulations on a classical computer.  
The algorithms we discuss here are all based on a second-quantized
formalism which naturally fits into the framework of quantum field
theory.  Another approach to simulating the many-body \Schrodinger
equation was discussed in \cite{compact}.

To begin the discussion, let us recall the form of the \Schrodinger
equation for a system of interacting particles moving in $d$
dimensions.

\subsection{Review of Schr\"odinger equation}
The Schr\"odinger
equation for a single free particle of mass $m$ moving in $d$ dimensions is
\begin{equation}
i \frac{\partial}{\partial t} \psi ({\bf x}, t) =
-\frac{1}{2m} \sum_{i} \frac{\partial^2}{\partial  (x^i)^2}
 \psi ({\bf x}, t)
\end{equation}
where  we have chosen units with the Planck constant  set to be 
$\hbar = 1$.  If the particle is moving in the presence of an external
potential $U (\bfx)$, the Schr\"odinger equation becomes
\begin{equation}
i \frac{\partial}{\partial t} \psi ({\bf x}, t) =
-\frac{1}{2m} \sum_{i}  \frac{\partial^2}{\partial  (x^i)^2}
\psi ({\bf x}, t)
+ U ({\bf x}) \psi ({\bf x}, t).
\end{equation}
For a system of $n$ identical particles interacting via a (symmetric)
pairwise potential $U ({\bf x}_i, {\bf x}_j)$, the Schr\"odinger
equation is
\begin{eqnarray}
i \frac{\partial}{\partial t} \psi ({\bf x}_1, \ldots, {\bf
x}_n, t) & = & \sum_{k = 1}^{n} 
-\frac{1}{2m} \sum_{i= 1}^{d}
 \frac{\partial^2}{\partial  (x_k^i)^2}
 \psi ({\bf x}_1, \ldots, {\bf x}_n, t) \label{eq:interacting}\\
 &  &\hspace*{0.3in}
 + \sum_{j< k} 
U ({\bf x}_j, {\bf x}_k) \psi ({\bf x_1}, \ldots, {\bf x_n}, t)
\nonumber
\end{eqnarray}

If we wish to simulate the $n$-particle Schr\"odinger equation in $d$
dimensions, one approach is to discretize space.  If we
discretize so that the particles move on a spatial lattice with $l$
lattice sites in each direction, the number of independent components
of the $n$-particle wavefunction grows as $l^{dn}$.  Even for $d = 3$,
for reasonably large values of $l$ and $n$ this number becomes
extremely large.  If we wish to simulate the system on a classical
computer, the number of independent components in the wavefunction is
a lower bound both on the amount of memory needed to store the state
of the system at a fixed point in time, and also for the amount of
computation needed to take a single time step in the simulation.  For
a lattice size of $l = 20$ and a system of $20$ particles, we have
$l^{3n} = 20^{60}\sim 10^{78}$, which is clearly far beyond the memory
and computational resources of any imaginable classical computer.  In
the following sections we will describe algorithms with which this
system can be simulated on a quantum computer with memory and
computational requirements per time step on the order of $l^d$, {\it
independent} of $n$ (so long as $n\ll l^d$).

\subsection{Simulating a free Schr\"odinger particle in one dimension}
We begin our discussion of simulations on a quantum computer by
considering the simplest case: a free Schr\"odinger particle moving on
a lattice in one dimension.  Let us consider a one-dimensional lattice
with $l$ vertices.  We will associate a single q-bit $\sigma_i$ with
each of the vertices $i$.  Let us associate a q-bit $\sigma_i$ in the
state $\uparrow$ with the presence of a ``particle'' at the $i$th
lattice site, and a q-bit in the state $\downarrow$ with the absence
of a particle.  The canonical basis for the Hilbert space contains
$2^l$ states, with anywhere between 0 and $l$ particles in the
available states.  Let us restrict attention to the single-particle
subspace of the Hilbert space.  This subspace has a basis containing
$l$ states; we denote by $|k\rangle$ the state where the single
particle is contained at lattice site $k$.  Graphically, this state
would look like
\begin{center}
\begin{picture}(200, 40)(- 100,- 25)
\multiput(- 60,0)( 20,0){3}{\makebox(0,0){$\downarrow$}}
\multiput(20,0)(20,0){3}{\makebox(0,0){$\downarrow$}}
\put(0,0){\makebox(0,0){$\uparrow$}}
\put(0,0){\circle*{4}}
\put(0,-15){\makebox(0,0){\scriptsize $k$}}
\put(20,-15){\makebox(0,0){\scriptsize $k + 1$}}
\put(-20,-15){\makebox(0,0){\scriptsize $k -1$}}
\put(45,-15){\makebox(0,0){$\cdots$}}
\put(-45,-15){\makebox(0,0){$\cdots$}}
\put(-85,0){\makebox(0,0){$\cdots$}}
\put(85,0){\makebox(0,0){$\cdots$}}
\end{picture}
\end{center}
We now wish to define a discrete time dynamics on the Hilbert space
which gives a \Schrodinger equation in the continuum limit.  Writing
the wave function describing the single-particle state at time $t$ in
the form
\begin{equation}
\psi (t) = \sum_{i}\psi_i (t) |i\rangle,
\end{equation}
the equation of motion for evolution over a single time step will be
\begin{equation}
\psi (t + 1) = M \cdot \psi (t)
\end{equation}
where $M$ is a unitary matrix.  One way of getting a
Schr\"odinger equation would be to find an operation on the full
Hilbert space which had the effect in the single particle Hilbert
space of acting by the operator
\begin{equation}
M = \left(\begin{array}{ccccccccc}
b & a & 0 &0 & \cdots & 0 & 0 & 0 & a\\
a &b & a & 0  & \cdots & 0 & 0 & 0 & 0\\
0 &a &b & a   & \cdots & 0 & 0 & 0 & 0\\
0 &0 &a &b   & \cdots & 0 & 0 & 0 & 0\\
& \vdots & & &\ddots & & & \vdots &\\
0 & 0 & 0 & 0 & \cdots & b & a & 0 & 0 \\
0 & 0 & 0 & 0 & \cdots & a &b & a & 0 \\
0 & 0 & 0 & 0 & \cdots & 0&a &b & a  \\
a & 0 & 0 & 0 & \cdots &0 & 0&a &b
\end{array} \right)
\label{eq:bogus}
\end{equation}
where $a, b$ are complex numbers satisfying $| b |^2 + 2 | a^2 | = 1$
and $a \bar{b} + \bar{a}b = 0$.  Unfortunately, this operator
cannot be implemented by a simple set of two q-bit operations
\cite{gz,Meyer}.  One
way of seeing that this is not possible is to take the inverse
of the matrix $M$.  The matrix $M^{-1}$ is a dense matrix for generic
lattice size $l$, indicating that $M$ cannot be implemented by
performing some sequence of local two q-bit operations at each lattice
site in a way which is independent of $l$.

It is worth noting at this point that although the operation $M$
cannot be simply realized in the framework of universal quantum
computers using q-bits and two q-bit operations, there are simple
physical systems which behave in a fashion very similar to the
dynamics which this operation would define.  In particular, a lattice
of spin-1/2 particles with a Hamiltonian  proportional to the
sum of exchange operations, as discussed in \cite{np}, would lead to a
very similar dynamics.  Thus, it may be advantageous to consider a
wider range of quantum computational devices in searching for
algorithms for simulating generic quantum systems.

Returning to the framework of universal quantum computers, we now
define a slightly more complicated dynamics which can be implemented
using two q-bit operations, and which leads to the desired Schr\"odinger
equation.  Consider the two operators acting on the single particle
Hilbert space
\begin{equation}
M_1 = \left(\begin{array}{ccccccccc}
b & a & 0 &0 & \cdots & 0 & 0 & 0 & 0\\
a &b & 0 & 0  & \cdots & 0 & 0 & 0 & 0\\
0 &0 &b & a   & \cdots & 0 & 0 & 0 & 0\\
0 &0 &a &b   & \cdots & 0 & 0 & 0 & 0\\
& \vdots & & &\ddots & & & \vdots &\\
0 & 0 & 0 & 0 & \cdots & b & a & 0 & 0 \\
0 & 0 & 0 & 0 & \cdots & a &b & 0 & 0 \\
0 & 0 & 0 & 0 & \cdots & 0&0 &b & a  \\
0 & 0 & 0 & 0 & \cdots &0 & 0&a &b
\end{array} \right)
\;\;\;\;\;
M_2 = \left(\begin{array}{ccccccccc}
b & 0 & 0 &0 & \cdots & 0 & 0 & 0 & a\\
0 &b & a & 0  & \cdots & 0 & 0 & 0 & 0\\
0 &a &b &  0   & \cdots & 0 & 0 & 0 & 0\\
0 &0 & 0 &b   & \cdots & 0 & 0 & 0 & 0\\
& \vdots & & &\ddots & & & \vdots &\\
0 & 0 & 0 & 0 & \cdots & b &  0 & 0 & 0 \\
0 & 0 & 0 & 0 & \cdots & 0 &b & a & 0 \\
0 & 0 & 0 & 0 & \cdots & 0&a &b &  0  \\
a & 0 & 0 & 0 & \cdots &0 & 0&0 &b
\end{array} \right)
\end{equation}
where  $| a |^2 + | b |^2 = 1$ and $a \bar{b} + b \bar{a} = 0$.
We can now describe a quantum algorithm which gives a time-evolution
equation for the single-particle sector
\begin{equation}
\psi (t) = M_1\cdot M_2 \cdot \psi (t-2)
\label{eq:dynamics}
\end{equation}
The essential point is that the single-particle state can be
transformed by either of the matrices $M_i$ by acting on the particles
in pairs by the two q-bit operator which is described in the basis $\{|
\downarrow \downarrow \, \rangle, | \uparrow \downarrow \, \rangle, |
\downarrow \uparrow \, \rangle, | \uparrow \uparrow \, \rangle\}$ by the
matrix
\begin{equation}
s =
\left(\begin{array}{cccc}
1 & 0 & 0 & 0\\
0 & b & a & 0\\
0 & a & b & 0\\
0 & 0 & 0 & 1
\end{array} \right)
\label{eq:single}
\end{equation}
Graphically, we can describe the implementation of the dynamics
(\ref{eq:dynamics}) through the following sequence of applications of
the operator $s$ to pairs of q-bits
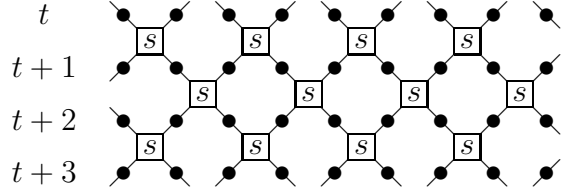
\begin{center}
\centering
\begin{picture}(200,90)(- 100,- 45)
\multiput(-80,30)(20,0){9}{\circle*{5}}
\multiput(-80,10)(20,0){9}{\circle*{5}}
\multiput(-80,-10)(20,0){9}{\circle*{5}}
\multiput(-80,-30)(20,0){9}{\circle*{5}}
\multiput(-70,20)(40,0){4}{\makebox(0,0){$s$}}
\multiput(-50,0)(40,0){4}{\makebox(0,0){$s$}}
\multiput(-70,-20)(40,0){4}{\makebox(0,0){$s$}}
\multiput(-75,25)(40,0){4}{\line(1,0){10}}
\multiput(-75,15)(40,0){4}{\line(1,0){10}}
\multiput(-75,25)(40,0){4}{\line(0, -1){10}}
\multiput(-65,25)(40,0){4}{\line(0, -1){10}}
\multiput(-55,5)(40,0){4}{\line(1,0){10}}
\multiput(-55,-5)(40,0){4}{\line(1,0){10}}
\multiput(-55,5)(40,0){4}{\line(0, -1){10}}
\multiput(-45,5)(40,0){4}{\line(0, -1){10}}
\multiput(-75,-15)(40,0){4}{\line(1,0){10}}
\multiput(-75,-25)(40,0){4}{\line(1,0){10}}
\multiput(-75,-15)(40,0){4}{\line(0, -1){10}}
\multiput(-65,-15)(40,0){4}{\line(0, -1){10}}
\multiput(-85,35)(40,0){5}{\line(1,-1){10}}
\multiput(-55,35)(40,0){4}{\line(-1,-1){10}}
\multiput(-65,15)(40,0){4}{\line(1,-1){10}}
\multiput(-75,15)(40,0){5}{\line(-1,-1){10}}
\multiput(-85,-5)(40,0){5}{\line(1,-1){10}}
\multiput(-55,-5)(40,0){4}{\line(-1,-1){10}}
\multiput(-65,-25)(40,0){4}{\line(1,-1){10}}
\multiput(-75,-25)(40,0){5}{\line(-1,-1){10}}

\put(-110, 30){\makebox(0,0){$t$}}
\put(-110, 10){\makebox(0,0){$t +1$}}
\put(-110, -10){\makebox(0,0){$t +2$}}
\put(-110, -30){\makebox(0,0){$t+ 3$}}

\end{picture}
\end{center}

We will now take the continuum limit of the dynamics
(\ref{eq:dynamics}) as the lattice spacing $\epsilon$ becomes small.
In the continuum limit,  we are interested in the behavior of
wavefunctions which are continuous and which therefore vary only by
order $\epsilon$ between adjacent lattice sites.
We can analyze the
dynamics of these wavefunctions in the continuum limit by performing a
mode analysis.  Assume that at time $t = 0$, we act with the matrix
$M_2$ which acts on the pairs of q-bits $(2j, 2j + 1)$.  If at time $t
= 0$ the even and odd lattice sites respectively have amplitudes given
by
\begin{eqnarray}
\psi_{2j}  (0)  & = & \alpha e^{i \kappa (2j)\epsilon} \\
\psi_{2j + 1}(0)  & = &  \beta e^{i \kappa (2j + 1)\epsilon} \nonumber
\end{eqnarray}
Then at time $t = 2$ we have
\begin{equation}
\left(\begin{array}{c}
\psi_{2j} (2) \\\psi_{2j + 1} (2)
\end{array}\right) =
\left(
\begin{array}{cc}
b^2 +  a^2 e^{-2i \kappa \epsilon} & ab (e^{i \kappa \epsilon} +e^{-i
\kappa \epsilon})\\ ab (e^{i \kappa \epsilon} +e^{-i \kappa
\epsilon})&  b^2 +  a^2 e^{2i \kappa \epsilon}
\end{array} \right) \cdot \left(\begin{array}{c}
\psi_{2j} (0) \\\psi_{2j + 1} (0)
\end{array} \right).
\label{eq:matrix1}
\end{equation}
For a fixed value of the angular wave number $\kappa$ the matrix which
appears in this equation has two eigenvalues.  In the continuum limit,
where $\kappa \epsilon \rightarrow 0$, the eigenvectors of this matrix
approach the limiting values $(1,1)$ and $(1, -1)$.  Since we are
interested in the behavior of wavefunctions which are smooth to order
$\epsilon$ at $t = 0$, only the first of these eigenvectors will be
relevant for describing time-development in the continuum limit.  The
eigenvalue associated with the first of these eigenvectors is
\begin{equation}
\lambda (\kappa \epsilon) =
b^2 +  a^2 \cos (2 \kappa \epsilon) +  a \cos (\kappa \epsilon)
\sqrt{4b^2 + 2 a^2 (\cos (2 \kappa \epsilon) -1)}
\end{equation}
Performing a power series expansion for small $\epsilon$ and fixed
$\kappa$ we have
\begin{equation}
\lambda (\kappa \epsilon) =
(a + b)^2 (1-\frac{a}{b}  \kappa^2 \epsilon^2 +{\cal O} (\epsilon^4))
\end{equation}
By factoring out an overall time-dependent phase from the wavefunction
\begin{equation}
\Psi (x, t) = (a + b)^{-t} \psi (x, t)
\end{equation}
we see that after scaling the time variable as $\epsilon^2$, the
wavefunction $\Psi$ obeys the dispersion relation
\begin{equation}
i \omega = \frac{a}{2b}  \kappa^2.
\label{eq:dispersion}
\end{equation}
Thus, in the continuum limit, to within an error of order $\epsilon$,
the wavefunction $\Psi (x, t)$ obeys the Schr\"odinger equation
\begin{equation}
i \partial_t \Psi (x, t) = -\frac{1}{2m}  \partial_x^2 \Psi (x, t)
\end{equation}
for a particle of mass
\begin{equation}
m = \frac{bi}{a} 
\label{eq:mass}
\end{equation}
Note that this mass is always real, due to the condition $a \bar{b} +
\bar{a}b = 0$ necessary for unitarity.

We have now seen that the dynamics (\ref{eq:dynamics}) gives rise in
the continuum limit to a Schr\"odinger equation for a single free
particle moving in one dimension.  A number of similar models have
been considered previously by other authors.  Essentially the same
model was studied by Succi and Benzi \cite{sb}; a number of numerical
experiments were performed on this model by Succi to study the
emergent Schr\"odinger behavior in the continuum limit \cite{Succi}.
A closely related model was described by Feynman \cite{fh} in the
context of the single particle Dirac equation in one spatial
dimension.  In Feynman's model, however, the amplitude $b$ was taken
to scale as the lattice spacing $\epsilon$.  This model was the basis
for the recent work of Meyer \cite{Meyer,Meyer2,Meyer3} on the
many-particle one-dimensional Dirac system; we will discuss this
system in Section 4.

\subsection{External potential}
We now have an algorithm for simulating a free quantum particle in one
dimension.  The dynamics of this system, which are described by the
dispersion relation (\ref{eq:dispersion}), are essentially 
trivial.  We shall now add various features to the model to give a
system of more physical interest.  First, we describe how an external
potential can be easily incorporated into the system.

Given an external potential function $U (x)$, the effect of this
potential can be brought into the dynamics by acting on each q-bit at
each time step with the matrix
\begin{equation}
U_x =\left(\begin{array}{cc}
1 & 0\\
0 &e^{-i \epsilon^2 U (x)}
\end{array} \right)
\label{eq:external}
\end{equation}
where $x$ is the position of the q-bit.  The inclusion of this
operation changes the time development of the system so that the
differential equation satisfied by $\Psi$ becomes
\begin{equation}
i \partial_t \Psi (x, t) = -\frac{1}{2m}  \partial_x^2 \Psi (x, t)
+ U (x) \Psi (x, t)
\end{equation}
in the continuum limit, precisely the Schr\"odinger equation for a
particle in an external potential.

It is worth noting that the continuum limit of the theory with an
external potential may not be well-behaved when the external potential
is not smooth.  If the potential is singular or has a discontinuity,
it is possible that the eigenvectors of the matrix in
(\ref{eq:matrix1}) with limiting form (1, -1) can be excited.  This
issue is discussed by Meyer in \cite{Meyer2}.  As long as the
potential function is smooth, however, the only eigenstates of the
system which are excited will be of the (1, 1) form, and the
Schr\"odinger equation in an external potential will be satisfied to
order $\epsilon$.  We have performed numerical analyses of the
spectrum for simple systems and have verified that for smooth
potentials the behavior of the model closely matches the continuum
system being modeled \cite{bw2}.

\subsection{Multiple particles}
Thus far, we have really not used the power of the quantum computer at
all.  The number of states available in the single-particle sector of
the Hilbert space is essentially the same as the number of q-bits, so
the algorithm for a single particle could just as well be implemented
on a classical computer.  Now, however, let us consider the
behavior of the multiple-particle sectors of the Hilbert space with
the same algorithm described above.  Keeping the notation that a q-bit
in the state $\uparrow$ corresponds to the presence of a particle, we
can for example  graphically describe a state with three particles at
positions $k-4, k, k + 2$ by
\begin{center}
\begin{picture}(300, 40)(- 150,- 25)

\multiput(- 120,0)( 20,0){2}{\makebox(0,0){$\downarrow$}}
\multiput(-60,0)(20,0){3}{\makebox(0,0){$\downarrow$}}
\multiput(20,0)(20,0){1}{\makebox(0,0){$\downarrow$}}
\multiput(60,0)(20,0){4}{\makebox(0,0){$\downarrow$}}

\put(-80,0){\makebox(0,0){$\uparrow$}}
\put(-80,0){\circle*{4}}
\put(0,0){\makebox(0,0){$\uparrow$}}
\put(0,0){\circle*{4}}
\put(40,0){\makebox(0,0){$\uparrow$}}
\put(40,0){\circle*{4}}
\put(0,-15){\makebox(0,0){\scriptsize $k$}}
\put(20,-15){\makebox(0,0){\scriptsize $k + 1$}}
\put(-20,-15){\makebox(0,0){\scriptsize $k -1$}}
\put(45,-15){\makebox(0,0){$\cdots$}}
\put(-45,-15){\makebox(0,0){$\cdots$}}
\put(-145,0){\makebox(0,0){$\cdots$}}
\put(145,0){\makebox(0,0){$\cdots$}}
\end{picture}
\end{center}
Because the operator $s$ in (\ref{eq:single}) preserves particle
number, the number of particles in the system is invariant under the
dynamics.  If we have a lattice of size $l$, as long as the number of
particles present $n$ is small compared to the lattice size ($n \ll
l$), generically particles will propagate for a fairly large number of
time steps without encountering other particles.  In the limit where
the lattice spacing gets arbitrarily small but the number of particles
remains fixed, we will obtain a \Schrodinger equation for $n$
particles, which will be noninteracting except when a pair of
particles are at the same lattice site.  Thus, we have described an
algorithm for simulating many Schr\"odinger particles in one dimension
interacting with a local ($\delta$ function) interaction potential.

By incorporating an external potential using single-particle operators
like (\ref{eq:external}) we can simulate a system of $n$ Schr\"odinger
particles moving in an external potential.  It is perhaps more
interesting, however, to include interactions between the particles
through a pairwise potential function $U (x, y)$.  A local interaction
corresponding to a potential $U (x, y) \sim \delta (x-y)$ with an
arbitrary phase can be incorporated by including a phase $\phi$ in the
two-particle component of the matrix $s$ in (\ref{eq:single}).
In general, however, we would like to incorporate nonlocal potential
functions.  In order to implement the effects of a general
interparticle potential, we can act at every time step on the
four-dimensional Hilbert space associated with every pair of q-bits at
positions $x$ and $y$ with the operator
\begin{equation}
U_{x, y} = \left(\begin{array}{cccc}
1 & 0 & 0 & 0\\
0 & 1 & 0 & 0\\
0 & 0 & 1 & 0\\
0 & 0 & 0 & e^{-i \epsilon^2 U (x, y)}
\end{array}\right).
\end{equation}
This operator acts on the state $\uparrow \uparrow$ with a nontrivial
phase, and therefore affects only the basis states where there are
particles at positions $x$ and $y$.  Just as in the case of the
external potential, incorporating the effects of these operators turns
the dynamical equation for $\Psi$ into a Schr\"odinger equation for
many interacting particles (\ref{eq:interacting}).

\subsection{Systems in arbitrary dimension $d$}

So far we have considered systems of particles moving only in a single
spatial dimension.  Of course,  most problems of physical interest involve
particles moving in three (or more) spatial directions.  We will now
discuss the generalization of the algorithms discussed above to higher
dimensions.  One simple way of simulating a  $d$-dimensional system
is to put the q-bits on a $d$-dimensional lattice, and to implement
the kinetic part of the time-development rule by repeatedly using the
algorithm described in section 3.2 along each of the dimensions
successively.  This leads to a fairly simple set of algorithms for
simulating the many-particle interacting Schr\"odinger equation in
$d$-dimensions.  There are several aspects of this algorithm, however,
which may be somewhat problematic.  For one thing, acting in each of
the directions in turn with the kinetic operator (\ref{eq:dynamics})
breaks the
underlying lattice symmetry.  For another thing, the number of q-bits
which behave differently under the time-development rule scales as
$2^d$, which for large $d$ may complicate the numerical interpretation
of simulations.

\begin{figure}
\begin{center}
\centering
\begin{picture}(200,100)(- 100,- 50)
\put(-45,19){\circle{5}}
\put(45,19){\circle{5}}
\put(-15,19){\circle*{5}}
\put(15,19){\circle*{5}}
\put(-45,-11){\circle{5}}
\put(45,-11){\circle*{5}}
\put(-15,-11){\circle{5}}
\put(15,-11){\circle{5}}
\put(-45,11){\circle*{5}}
\put(45,11){\circle*{5}}
\put(-15,11){\circle{5}}
\put(15,11){\circle{5}}
\put(-45,-19){\circle{5}}
\put(45,-19){\circle*{5}}
\put(-15,-19){\circle{5}}
\put(15,-19){\circle*{5}}

\put(-49,15){\circle{5}}
\put(49,15){\circle{5}}
\put(-19,15){\circle{5}}
\put(19,15){\circle*{5}}
\put(-49,-15){\circle{5}}
\put(49,-15){\circle{5}}
\put(-19,-15){\circle{5}}
\put(19,-15){\circle{5}}
\put(-41,15){\circle*{5}}
\put(41,15){\circle{5}}
\put(-11,15){\circle*{5}}
\put(11,15){\circle*{5}}
\put(-41,-15){\circle*{5}}
\put(41,-15){\circle{5}}
\put(-11,-15){\circle*{5}}
\put(11,-15){\circle{5}}

\multiput(- 38, 15)(30,0){4}{\vector(1,0){16}}
\multiput(-38, -15)(30,0){4}{\vector(1,0){16}}
\multiput(- 52, 15)(30,0){4}{\vector(-1,0){16}}
\multiput(-52, -15)(30,0){4}{\vector(-1,0){16}}

\multiput(-45,8)(30,0){4}{\vector(0,-1){16}}
\multiput(-45, -22)(30,0){4}{\vector(0,-1){16}}
\multiput(-45,-8)(30,0){4}{\vector(0,1){16}}
\multiput(-45, 22)(30,0){4}{\vector(0,1){16}}

\end{picture}
\end{center}
\caption[x]{\footnotesize A typical basis state in the Hilbert space
of a 2D QLGA.  Filled circles correspond to occupied particle states
(q-bits in the state $\uparrow$).  Arrows correspond to directions of
propagation.}
\label{f:state}
\end{figure}
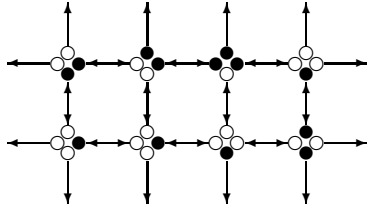

A nice way of packaging the q-bits and kinetic time-development rule
in $d$-dimensions is given by the general class of systems called
Quantum Lattice-Gas Automata (QLGA) \cite{Meyer,bw1}.  These models are
defined by analogy with classical lattice-gas automata \cite{lga} to be
simple dynamical systems in which q-bits on a lattice undergo a
time-development which consists of alternating propagation and collision
steps.  At each lattice site there are a set of q-bits corresponding
to particle occupation sites for particles with velocity vectors lying
on the lattice.  On a cartesian lattice in $d$-dimensions, for
example, we could have $2d$ q-bits at each lattice site, corresponding
to sites for particles moving along each of the basis vectors.  Just
as in the previous models, we associate a q-bit in state $\uparrow$
($\downarrow$) with the presence (absence) of a particle.  
These models have the advantage that they can be defined in a fashion
which is completely symmetric with respect to the discrete symmetry
group of the underlying lattice.  Also, the number of distinct types
of q-bits goes as $2d$ instead of $2^d$ as in the iterative model
mentioned above.
An example of a basis state for the Hilbert space of a QLGA in two
dimensions is given in Figure~\ref{f:state}.  The filled circles
represent occupied particle sites, and the empty circles represent
unoccupied sites.  

In the propagation stage of the time-development of a
QLGA, the state of the system is transformed by a permutation on the
q-bits.  This permutation takes the q-bit associated with each
position and velocity to a q-bit at the lattice site reached by adding
the velocity vector to the position.  As indicated by the arrows in
the example in
Figure~\ref{f:state}, this transformation can be achieved by
exchanging the q-bit states in pairs.  For each pair of q-bits which
must be exchanged, the operator acting on the Hilbert space is given
in the basis $\{| \downarrow \downarrow \, \rangle, | \uparrow \downarrow
\, \rangle, | \downarrow \uparrow \, \rangle, | \uparrow \uparrow \,
\rangle\}$ by the matrix
\begin{equation}
E =
\left(\begin{array}{cccc}
1 & 0 & 0 & 0\\
0 &  0 & 1 & 0\\
0 & 1 & 0 & 0\\
0 & 0 & 0 &  1
\end{array} \right)
\end{equation}

In the collision stage of the time-development, the state associated
with all the q-bits at each lattice site is acted on by a unitary
``collision'' operator.  This operator is chosen to conserve particle
number.  In general, we would like to choose a collision operator
which is invariant under the group of discrete symmetries of the
lattice.  As in the one-dimensional model discussed above, for systems
where the particle number is fixed and the lattice spacing becomes
small, the part of the collision operator in the single-particle
sector determines the propagation of the individual particles.  On a
cartesian lattice in $d$ dimensions, the collision operator in the
single particle sector is parameterized by 3 complex phases $\mu, \nu,
\lambda$.  These phases correspond to the eigenvalues of the collision
matrix for vectors which lie in the three irreducible representations
of the discrete rotation group into which the $2 d$-dimensional single
particle space decomposes.  Specifically, $\mu$ is the eigenvalue
associated with the constant vector $(1, 1, \ldots, 1)$ and $\nu$ is
the eigenvalue of vectors which are antisymmetric under a parity
transformation.  It is shown in  \cite{bw1} that in the generic case
where the three eigenvalues are distinct, the behavior of the
individual particles in the system is governed by a Schr\"odinger
equation for a nonrelativistic particle with mass $m$ given by the relation
\begin{equation}
\frac{i}{2m}  = \frac{1}{d}  \left(\frac{\nu}{\mu-\nu}  +\frac{1}{2}\right).
\label{eq:mass2}
\end{equation}

By combining the propagation and collision processes into a single
time-development rule, we have a QLGA model which has a continuum
limit described by a many-particle Schr\"odinger equation in $d$
dimensions.  As described so far, the particles in this system interact
only through local ($\delta$ function) interactions, which are
parameterized by the part of the collision matrix acting in the
multiple-particle space.  Just as in the one-dimensional model
discussed above, it is straightforward to include external and
interparticle potentials, so that an arbitrary system of
nonrelativistic interacting quantum particles can be simulated using
this type of model.

In fact, the one-dimensional model discussed in sections 3.2-3.4 is
precisely equivalent to a simple one-dimensional QLGA.  We can
associate the even 
and odd lattice sites with right- and left-moving particle sites at
even time steps, and with left- and right-moving particle sites at odd
time steps.  If we define a one-dimensional QLGA with a collision
matrix given by
$s$ as in (\ref{eq:single}), it is easy to see that the QLGA
decomposes into two noninteracting copies of the one-dimensional model
defined through (\ref{eq:dynamics}).  For this choice of collision
matrix we have $\mu = a + b$ and $\nu = a-b$ (the sign on $\nu$ arises
because the collision operator $s$ implicitly changes the direction of
velocity for each particle).  Inserting these values for $\mu, \nu$ into
(\ref{eq:mass2}) reproduces (\ref{eq:mass}) as we would expect.

It should be noted that the algorithms we have discussed here are most
easily used to simulate systems of interacting bosons.  The
microscopic quantum system only contains a single state in the Hilbert
space corresponding to each configuration of particles.  Two paths in
which particles move to positions $x, y$ and $y, x$ will contribute
amplitudes which add with no change of sign, as is appropriate for
Bose statistics.  It is possible to modify the algorithm to
incorporate Fermi statistics, however this requires somewhat more
bookkeeping.  This issue was discussed in  \cite{al}.

\subsection{Computational complexity}

We have now described a class of algorithms for simulating the
nonrelativistic many-body Schr\"odinger equation on a quantum
computer.  As discussed in Section 3.1, simulating $n$ quantum
particles on a lattice of size $l^d$ requires a computation of
complexity ${\cal O} (l^{dn})$ on a classical computer.  We now
briefly discuss the time and memory requirements for such a simulation
on a quantum computer.  Using the quantum lattice-gas automaton
formalism, we can simulate such a system with  $2d
\cdot l^d$ q-bits.  The propagation step takes on the order of $dl^d$ two
q-bit operations, since the q-bits are simply exchanged pairwise.  The
free kinetic collision step requires on the order of $d^2 l^d$
operations (the factor of $d^2$ is needed to construct the collision
operator at each vertex from two q-bit operations).  Incorporating an
arbitrary interparticle interaction potential would require on the
order of $d^2 l^{2d}$ operations, since we must act on each pair of
q-bits independently.  This operation will clearly be the most
computationally expensive part of the simulation.  
We see, however,
that the memory and time requirements for the algorithms we have
discussed are only polynomial in the lattice size, and are completely
independent of the number of particles as long as $n\ll l^d$.  Thus,
we can achieve a speedup for many-particle quantum simulations which
is exponential in the number of particles being simulated.  For
example, for a lattice of size $l = 20$ in three spatial dimensions,
the number of operations needed in a simulation would be on the order
of $36\cdot 20^6\sim 2.3\cdot 10^9$ compared with $10^{78}$ operations for a
simulation on a classical computer.

In discussing the computational complexity of these simulation
algorithms, it is worth noting that if we simulate nonrelativistic
fermions extra work must be done to keep track of the phase of the
state.  Using the bookkeeping method suggested in \cite{al}, for
example, we must define a canonical ordering for all the particles,
and we must check to see when the propagation changes the ordering of
any pair of particles.  This will take roughly on the order of $d^2
l^{2d}$ operations per time step, just as for the interparticle
interaction.  In fact, however, we can incorporate this phase into the
interparticle interaction matrix, so that the simulation of an
interacting system of nonrelativistic fermions can be achieved in
precisely the same number of quantum operations as the simulation of
interacting ``hard'' bosons (particles with bosonic statistics which
cannot occupy the same lattice site).

\section{Simulating relativistic systems}

In the bulk of this paper we have discussed the simulation of
nonrelativistic systems of many Schr\"odinger particles.  Many systems
of physical interest are described to a high degree of precision by the
nonrelativistic many-body Schr\"odinger equation.  However, there are
many physical systems  for which the Schr\"odinger equation is not a
sufficiently accurate model.  In particular, it is desirable to have a
way of simulating relativistic quantum systems such as gauge fields or
Dirac fermions.

Let us first consider the problem of simulating free Dirac fermions in
$d$ dimensions on a quantum computer.  As mentioned in Section 3.2, it
was found by Feynman
 \cite{fh} that if the parameter $b$ is taken to approach 0 at the same
rate as the lattice spacing, the one-dimensional model described above
satisfies the first-order two-component Dirac equation.  By
generalizing this single-particle model to the QLGA framework, it was
shown by Meyer  \cite{Meyer} that a free system of many noninteracting
Dirac particles in 1 + 1 dimensions could be simulated on a quantum
computer.

The problem of simulating Dirac fermions on a lattice in more than one
spatial dimension has presented difficulties to physicists for decades
now.  A full discussion of the approaches which have been used to try
to solve this problem is beyond the scope of this paper.  The
best-known obstacle to a sensible discretization of a field theory of
Dirac particles is the fermion doubling problem.  If we try naively to
generalize Feynman's model for the one-dimensional Dirac system to
higher dimensions, we immediately encounter several difficulties.
First, even for a single particle it is not straightforward to find a
lattice rule which gives the propagator for the Dirac equation.
Within the QLGA framework, in fact, it is not possible to construct a
single-particle Dirac equation in any dimension $d > 1$.  A model
suggested several years ago by Bialynicki-Birula \cite{bb}, however,
fits very closely into the framework we have been discussing, and may
provide a way of simulating systems of many Dirac particles without
encountering the difficulties which have plagued previous efforts.

The basic approach of Bialynicki-Birula is to construct a quantum
cellular automaton model with a single time-development rule which
depends only on local variables.  He finds that in three spatial
dimensions, on the body-centered cubic lattice where every vertex has
eight neighbors, it is possible to find such a rule whose continuous limit
corresponds to a two-component Weyl spinor.  Remarkably, it seems that
such a model cannot exist on the three-dimensional cartesian lattice.  A Dirac
particle is formed by taking two Weyl spinors and including a mass
term by hand.  As described by Bialynicki-Birula, the model avoids the
no-go theorem of Nielsen and Ninomiya \cite{nn} because it has
discrete time evolution.  The model of Bialynicki-Birula suffers from
the same difficulty as the model with time-development operator
(\ref{eq:bogus}) which we discussed in Section 3.2, and thus this
algorithm cannot be implemented naturally on a universal quantum
computer.  However, the existence of the model suggests that by using
a multi-step rule on the body-centered cubic lattice, it might be
possible to efficiently simulate free Dirac fermions in three
dimensions on a quantum computer.

If one could indeed simulate many free Dirac particles on the lattice,
it is natural to try to simulate a full second-quantized field theory
of Dirac particles.  To do this, one would need to introduce
antiparticles  corresponding to negative energy states.
Incorporating a potential directly into a Dirac theory is liable to be
problematic, however, since the simplest interaction term, the four-fermi
interaction, is nonrenormalizable and would destroy the continuum
limit of the theory  \cite{Susskind}.

Of course, if one wishes to simulate interacting Dirac fermions, one
is led naturally to the problem of simulating quantum gauge field
theories.  Finding an algorithm with which one could simulate a
quantum field theory of interacting gauge fields and fermions on a
quantum computer with exponential performance enhancement would be a
very interesting result.  The QLGA approach is naturally suited to
describing second-quantized systems in a Hamiltonian framework, so
such a model is probably best discussed in the context of
Hamiltonian lattice methods  \cite{ks}.  

There are several difficulties inherent in constructing a QLGA model
of a quantum gauge theory.  First, it is necessary to find a
lattice rule which describes the propagation of a single particle
according to Maxwell's equations.  This is a challenge, because it is
difficult to have an isotropic propagator on a lattice when the
scaling of the spatial and temporal lattice spacing is the same.
Bialynicki-Birula has suggested that his model for the Weyl spinor can
be used to describe a propagating photon by writing the Maxwell
equations in spinor notation.  It is unclear how to formulate this
model in QLGA language, however it does not seem that there is any
fundamental obstacle to such a formulation.  Another general
difficulty with simulating gauge theories is the discretization of the
gauge group.  In the QLGA formalism, however, where we are essentially
dealing with a second-quantized field theory, this should not be a
problem; the fermions will carry a gauge index, and the gauge
particles will carry an index in the adjoint representation of the
gauge group.  Thus, for example, to simulate the theory of QCD with an
SU(3) gauge theory, we would include three colors of fermions and 8
species of gluons, just as in the physical particle theory we are
simulating.

To end this section, we briefly discuss a completely different
approach to simulating quantum field theories on a quantum computer
which does not rely on the QLGA framework.  It is always possible to
describe a field theory with a momentum cutoff by writing the
Hamiltonian in terms of creation and annihilation operators for the
modes of the fields with momentum below the cutoff.  In a discrete
time formulation, the evolution of the state of the system in a single
time step is given by the exponential of the Hamiltonian times the
time step (times $i$).  If this time-development operator can be
reasonably approximated by a polynomial number of two-particle
interactions, then such a system could be simulated on a quantum
computer with an exponential performance increase using the methods
discussed in \cite{Lloyd}.  Whether such a nonlocal approach to the
simulation is preferable to the local QLGA method is not clear;
further study of both approaches will probably be needed to resolve
this issue.

\section{Conclusions}

We have discussed algorithms for simulating quantum systems on quantum
computers.  In particular, we have described in detail algorithms
which enable the simulation of interacting many-body quantum systems
with exponential performance enhancement over classical computers.
There are many obstacles to building the sort of large-scale quantum
computers which would be necessary for the algorithms discussed here
to be useful.  In addition to the technical difficulties involved with
precisely controlling systems of many quantum elements, there are the
theoretical problems of making algorithms robust against imprecision
and loss of coherence which must appear in any physical system.  Many
of the other speakers at this conference have discussed these and
related issues in great detail, so we have concentrated here on
describing algorithms under the assumption that eventually the
technical problems will be solved and we will have a programmable
universal quantum computer at our disposal.  Certainly, recent
progress in quantum error correcting codes \cite{codes} gives some
indication that there is room for optimism in this direction.  Even if
a true ``universal quantum computer'' can never be built, however, it
may be possible to construct quantum devices which have some
computational capacity \cite{others}.  Because the algorithms we have
discussed here are based only on simple local quantum interactions, it
seems plausible that many different types of quantum systems may be
useful in simulating other quantum systems in a precise and relatively
controlled fashion.  

One interesting aspect of algorithms which simulate quantum systems on
a quantum computer is the issue of measurement.  In a classical
computation, all the bits in the system are available at all times to
the programmer, so that the state of the system can be completely
known.  In a quantum computer, on the other hand, the state of the
system can only be known through the measurement of certain q-bits,
whose values are determined probabilistically through the usual rules
of quantum mechanics.  This means that some information in the state
of a quantum computer is not directly accessible to the programmer.
In terms of the algorithms discussed here for simulating quantum
systems, this means that only certain questions can be asked of the
simulation.  For example, imagine simulating a system of a single
electron in an external potential well.  In principle, given the exact
time-development of the wavefunction in the quantum computer doing the
simulation, a spectral analysis would immediately enable one to
calculate the entire energy spectrum of the system.  However, this
complete information is not available.  Instead, we can only ask the
types of questions which we could ask in the real physical system--for
example, we could ask for the probability that a particle which starts
at a fixed point in space moves to a given region in a fixed time.
This limitation on the set of questions which can be meaningfully
asked of a quantum computer may seem at first to make the
process of extracting meaningful results from such simulations  more
difficult.  However, since the questions which can be asked of the
simulation are precisely those questions which can be asked of the
system being simulated, the connection to results of physical
experiments is very direct.  Any difficulty in connecting the results
of the simulation with predictions about physical systems must be the
consequence of theoretical prejudices.

\section*{Acknowledgements}
We would like to acknowledge helpful conversations with Dan Abrams,
Seth Lloyd, Vipul Periwal and Jeffrey Yepez.  We would like to thank
the organizers of the PhysComp '96 conference for providing a
stimulating environment, and Boston University for their hospitality
during the conference.  BMB was supported in part by an IPA from
Phillips Laboratories and in part by the United States Air Force
Office of Scientific Research under grant number F49620-95-1-0285.  WT
was supported by the National Science Foundation (NSF) under contract
PHY96-00258.

\end{document}